\documentclass[]{emulateapj-rtx4}

\usepackage{natbib,color} 
\bibpunct{(}{)}{;}{a}{}{,}

\DeclareRobustCommand{\ion}[2]{%
\relax\ifmmode
\ifx\testbx\f@series
{\mathbf{#1\,\mathsc{#2}}}\else
{\mathrm{#1\,\mathsc{#2}}}\fi
\else\textup{#1\,{\mdseries\textsc{#2}}}%
\fi}

\shorttitle{Fast inversion of solar \ion{Ca}{ii} spectra}
\shortauthors{Beck, C.; Choudhary, D.P.; Rezaei, R.; Louis, R.E.}

\begin{document}
\title{Fast inversion of solar \ion{Ca}{ii} spectra} 
%  with a non-local thermodynamic equilibrium temperature correction


\author{C. Beck}
\affil{National Solar Observatory (NSO)}
\author{D.P. Choudhary}
\affil{Department of Physics \& Astronomy, California State University, Northridge (CSUN)}
\author{R. Rezaei}
\affil{Kiepenheuer-Institut f\"ur Sonnenphysik (KIS)}
\author{R.E. Louis}
\affil{Leibniz-Institut f\"ur Astrophysik (AIP)}


%\altaffiltext{1}{Departamento de Astrof\'isica, Universidad de La Laguna, E-38205 La Laguna, Tenerife, Spain}


\begin{abstract}
We present a fast ($\ll 1$\,s per profile) inversion code for solar \ion{Ca}{ii} lines. The code uses an archive of spectra that are synthesized prior to the inversion under the assumption of local thermodynamic equilibrium (LTE). We show that it can be successfully applied to spectrograph data or more sparsely sampled spectra from two-dimensional spectrometers. From a comparison to a non-LTE inversion of the same set of spectra, we derive a first-order non-LTE correction to the temperature stratifications derived in the LTE approach. The correction factor is close to unity up to $\log\,\tau\sim -3$ and increases to values of 2.5 and 4 at $\log\,\tau = -6$ in the quiet Sun and the umbra, respectively.
\end{abstract}

\section{Introduction}
One of the main problems in the analysis of observational data in solar physics is how to extract the physical properties of the solar atmosphere. These can be either the thermodynamic state of the atmosphere, namely, temperature, density, mass motions or the number of all available degrees of freedom, or the vector magnetic field. For spectroscopic and spectropolarimetric observations of the solar photosphere, several methods with a differing degree of sophistication exist. These range from simple proxies such as the location of line cores of spectral lines for a determination of the velocity to complex approaches such as a detailed modeling of the solar atmosphere. The latter approach is the backbone of so-called inversion codes that do not try to derive information directly from the observations, but instead use a model of the solar atmosphere to reproduce the observations.

For solar photospheric spectra several inversion codes are available. The most basic ones use the Milne-Eddington (ME) approximation that provides an analytic solution of the radiative transfer equation \citep{skumanich+lites1987,socasnavarro+etal2001,lites+etal2007,borrero+etal2011,borrero+etal2014}. The ME approximation can be improved by employing the condition of local thermodynamic equilibrium (LTE) instead, which is used in, e.g., the Stokes Inversion based on Response functions code \citep[SIR;][]{ruizcobo+toroiniesta1992} or the Stokes-Profiles INversion-O-Routines \citep[SPINOR;][]{frutiger+etal2000,berdyugina+etal2003}. Additional improvements and/or modification to such LTE inversion codes were a more rigid approach to find the best fit from an absolute minimum in the $\chi^2$ surface using Bayesian methods \citep{asensioramos+etal2007,asensioramos+etal2012} or the inclusion of spatially distributed information \citep{vannoort2012}.

The LTE condition breaks down in the upper solar photosphere and in the solar chromosphere. Compared with the available inversion codes for photospheric data, there are fewer chromospheric non-LTE inversion codes, with often also an additional restriction to specific spectral lines \citep[e.g.,][]{tziotziou+etal2001,teplitskaya+grigoryeva2009}. The most general approach is used in the  Non-LTE Inversion COde using the Lorien Engine \citep[NICOLE;][]{socasnavarro+etal2000,socasnavarro+etal2014} that was recently released\footnote{https://github.com/hsocasnavarro/NICOLE/}. The HAnle and ZEeman Light \citep[HAZEL;][]{asensio+etal2008} inversion code is more aimed at the determination of magnetic field properties for specific spectral lines. 

One major drawback of more sophisticated inversion approaches that include more realistic physics (ME--LTE--NLTE) is the required computational effort. A ME inversion of several thousand profiles can be done in a few minutes, LTE codes commonly need a few seconds per profile to be inverted, while NLTE codes can require up to a few minutes per profile. This has hindered the regular application of inversion codes to most chromospheric observations. Near real-time results on the chromospheric thermal structure and its evolution, e.g., in active regions could be helpful for improving flare prediction capabilities. Here, we present a fast inversion code for solar chromospheric \ion{Ca}{ii} lines that could be extended to other lines as well. We note that for the hydrogen Balmer H$_\alpha$ line, no corresponding analysis tools are available to date.
\section{Observations}
\begin{figure}
\centerline{\resizebox{8.5cm}{!}{\includegraphics{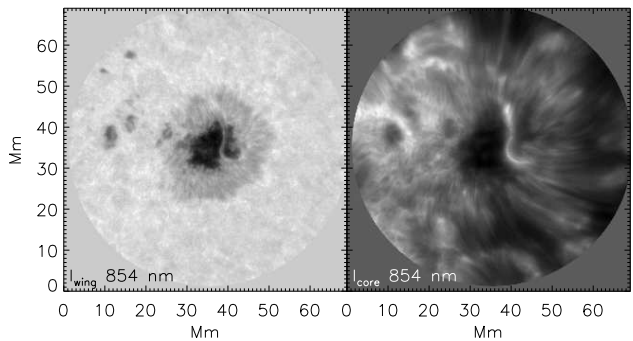}}}
\caption{Overview of the FOV. Left/right panel: line-wing and line-core image of one spectral scan of Ca 854.2\,nm with IBIS.\label{fov}}
\end{figure}

We observed the leading spot of NOAA 12002 on March 13, 2014, from about UT 20:37 to UT 21:00 simultaneously with the Interferometric BI-dimensional Spectrometer \citep[IBIS, imaging spectrometer;][]{cavallini2006,reardon+cavallini2008} and the SPectropolarimeter for Infrared and Optical Regions \citep[SPINOR, slit spectrograph;][]{socasnavarro+etal2006}. The sunspot was located at $x, y \sim 40^{\prime\prime}, -180^{\prime\prime}$ at a heliocentric angle of about 12 degrees. Figure \ref{fov} shows an overview image of the sunspot. It harbored a distinct light bridge (LB) that separated its umbra into two umbral cores.

With IBIS, we sampled the chromospheric spectral lines of \ion{Ca}{ii} IR at 854.2\,nm (in the following Ca 854.2\,nm) and H$_\alpha$ at 656\,nm. IBIS used an exposure time of 80\,ms and sequentially scanned Ca 854.2\,nm and H$_\alpha$ with a non-equidistant sampling of 27 wavelength points for each line. The spectral sampling was about 4 (40)\,pm pixel$^{-1}$ at the line core (wing) covering a range from 853.93 to 854.48\,nm. The spatial sampling was 0\farcs098 inside a circular field of view (FOV) of about 70\,Mm diameter. IBIS obtained 63 individual scans through the spectral lines during the observing sequence at a cadence of about 13\,s and under mediocre seeing conditions.

With SPINOR, we obtained Stokes vector polarimetry of \ion{He}{i} at 1083\,nm and \ion{Ca}{ii} IR at 854.2\,nm, and spectroscopic data of H$_\alpha$. We scanned the FOV with 200 steps of 0\farcs3 step width. The exposure time per modulation state in SPINOR was 100\,ms and the signal was integrated over about 5\,s at each step. The spatial sampling along the slit was 0\farcs366 and the spectral sampling was 5.85\,pm pixel$^{-1}$ for Ca 854.2\,nm. The full wavelength window covered a range from 852.92 to 855.90\,nm. Individual spectra from both instruments are displayed in Figures \ref{single_spec} and \ref{IBIS_single_spec}.
\section{Inversion of Ca 854.2\,nm spectra}
\subsection{Assuming Local Thermodynamic Equilibrium}
All spectroscopic data in Ca 854.2\,nm from either SPINOR or IBIS were inverted with a modified version of the inversion code for \ion{Ca}{ii} lines that assumes LTE and that is described in detail in \citet{beck+etal2013,beck+etal2013a} and \citet{beck+etal2014}. We improved the generation of the pre-calculated spectral archive by enforcing hydro-statical equilibrium of the perturbed temperature stratifications prior to the spectral synthesis. This was done with the program ``optical'' that is part of the SIR code. The LTE inversion code finds the best match to an observed profile of a \ion{Ca}{ii} line by a comparison to all of the about 240,000 archive spectra and subsequently retrieves the corresponding temperature stratification. \citet{beck+etal2013a}, \citet{beck+etal2014} and the Appendix discuss the accuracy that can be achieved in the reproduction of observed profiles.

For the SPINOR spectra, about 180 wavelength points were overlapping with those of the archive spectra. The IDL code required about 0.8\,s per profile on a quad-core 2.4 GHz machine with 8 GB RAM and Ubuntu/Linux as operating system. For IBIS data with 27 wavelength points, this reduced to about 0.1\,s per profile on the same computer. Running the code on a machine with 32 cores of 2.7\,GHz and 64 GB of RAM with Solaris as operating system, the time per profile reduced to 0.07\,s per profile when running a single job and about 0.2\,s per profile for ten jobs in parallel. The IDL code was optimized for speed by making extensive use of the IDL-implemented matrix and vector operations that are executed outside of IDL and that thus profit from a multi-core architecture even if the corresponding IDL session itself uses only a single core. A single spectral scan of IBIS contains about 700,000 individual profiles inside the circular aperture and therefore required about one day of computing time using one job.
 
The archive of pre-calculated \ion{Ca}{ii} spectra, the corresponding temperature and electron density stratifications, one example IBIS data set and the necessary set of IDL routines can be downloaded from ftp.nso.edu:/outgoing/cbeck/ca\_inv\_exam/. Please check the README.TXT in the folder for more details on the usage.
\begin{figure*}
\centerline{\resizebox{17.cm}{!}{\includegraphics{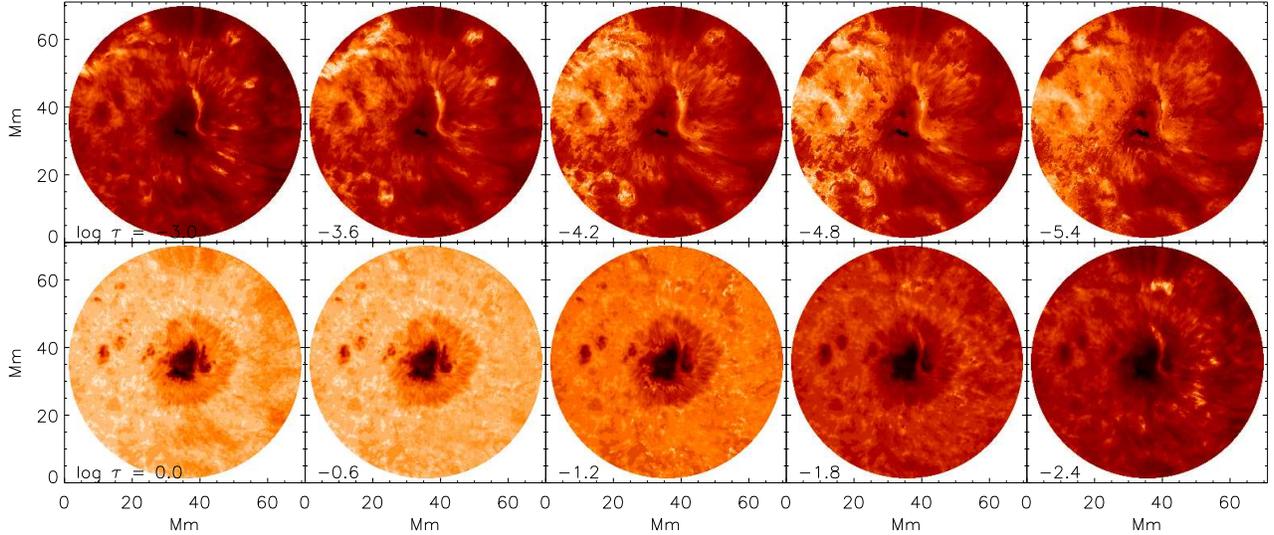}}}
\caption{Temperature maps at different levels of $\log \tau$ in the inversion of one spectral scan of Ca 854.2\,nm with IBIS. See Figure \ref{nlte_lte} below for the approximate temperature ranges.\label{taumaps_ibis}}
\end{figure*}
\begin{figure}
\centerline{\resizebox{8.5cm}{!}{\includegraphics{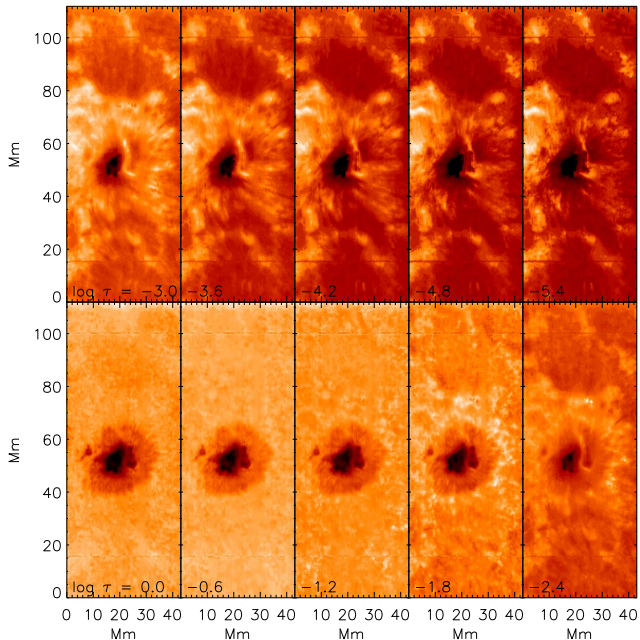}}}
\caption{Temperature maps at different levels of $\log \tau$ in the inversion of the SPINOR Ca 854.2\,nm data. See Figure \ref{nlte_lte} below for the approximate temperature ranges. \label{taumaps_spinor}} %All images are displayed in their full dynamical range; s
\end{figure}
\subsection{Assuming Non-Local Thermodynamic Equilibrium}
The polarimetric data in Ca 854.2\,nm obtained by SPINOR were also inverted with the NICOLE code \citep[see also][]{delacruzrodriguez+etal2012,delacruzrodriguez+etal2013a}. We assumed a single atmospheric component and used three cycles in each inversion. We repeated the inversion process with a varied initial model ten times for each pixel and retained the best result from the ten inversions. We allowed for two, three and five nodes in temperature in the first, second and third cycle, respectively. The corresponding numbers of nodes for the line-of-sight (LOS) velocity were one, two, and three nodes. The properties of the magnetic field, microturbulence, macroturbulence and the stray-light fraction were assumed to be height-independent in all cycles. The initial model atmosphere outside the LB area was very similar to the Harvard Smithsonian Reference Atmosphere \citep[HSRA;][]{gingerich+etal1971} model. 

For all pixels in the LB and its surroundings, we started five inversions in which we prescribed velocity peaks with a Gaussian profile at a certain height in the atmosphere because of their complex shape (cf.~Figures \ref{single_spec} and \ref{IBIS_single_spec} below). These velocity profiles had peak values from -7 to +28 km\,s$^{-1}$. Although we allowed the initial velocity stratification to correspond to a blue shift, a majority of the best-fit NLTE profiles that successfully reproduced the complex shape of the observed Ca 854.2\,nm spectra were finally red-shifted, in agreement with spatial maps of the LOS line-core velocity of that line. 

One major drawback of the complete solution of the NLTE problem is the required computing time. The NICOLE code needed about 5 min per profile (depending on the optimization and regularization scheme) on SPINOR spectra with about 400 wavelength points when running a single pixel on a single core. The NLTE inversion results turned out to be noisier in the spatial variation of inversion parameters despite using a very limited number of degrees of freedom. We thus run a 3$\,\times\,$3 spatial smoothing over the corresponding two-dimensional (2D) maps (cf.~Figure \ref{nlte_lte} below). Note that the stratification of physical parameters in NLTE inversions spans a larger optical depth range (up to $\log \tau = -8$) compared to the LTE inversion (up to $\log \tau = -6$) as seen in Fig.~\ref{nlte_corr}. 

The appendix shows examples of the fit quality of the LTE fit to the SPINOR and IBIS data and of the NLTE fit to the SPINOR data.
\begin{figure}
\centering
\resizebox{4.3cm}{!}{\includegraphics{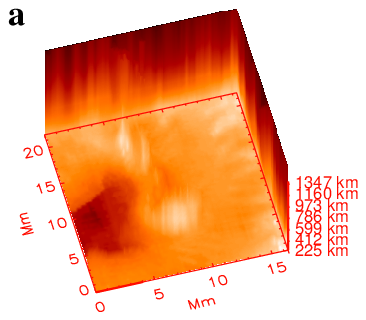}}\resizebox{4.3cm}{!}{\includegraphics{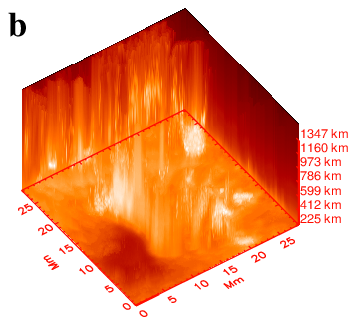}}\\
\resizebox{4.3cm}{!}{\includegraphics{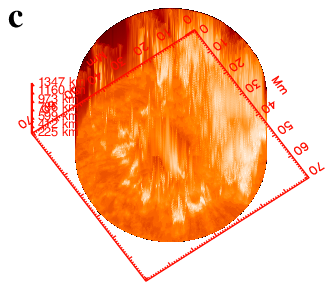}}\resizebox{4.3cm}{!}{\includegraphics{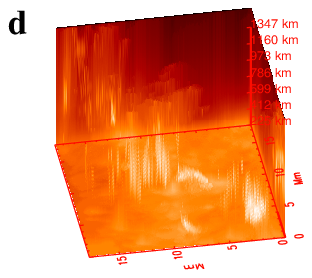}}\\
\caption{3D temperature renderings. Panels a/b: LB region from the inversion of the SPINOR/IBIS Ca 854.2\,nm spectra. Panels c/d: full FOV and  loop structure in the moat from the inversion of one spectral scan of Ca 854.2\,nm with IBIS. See also animation 1 ({\bf a}), animation 2 ({\bf c}) and animation 3 ({\bf d}) in the on-line section.\label{3d_spinor}}
\end{figure}

\begin{figure*}
\centerline{\resizebox{17.cm}{!}{\includegraphics{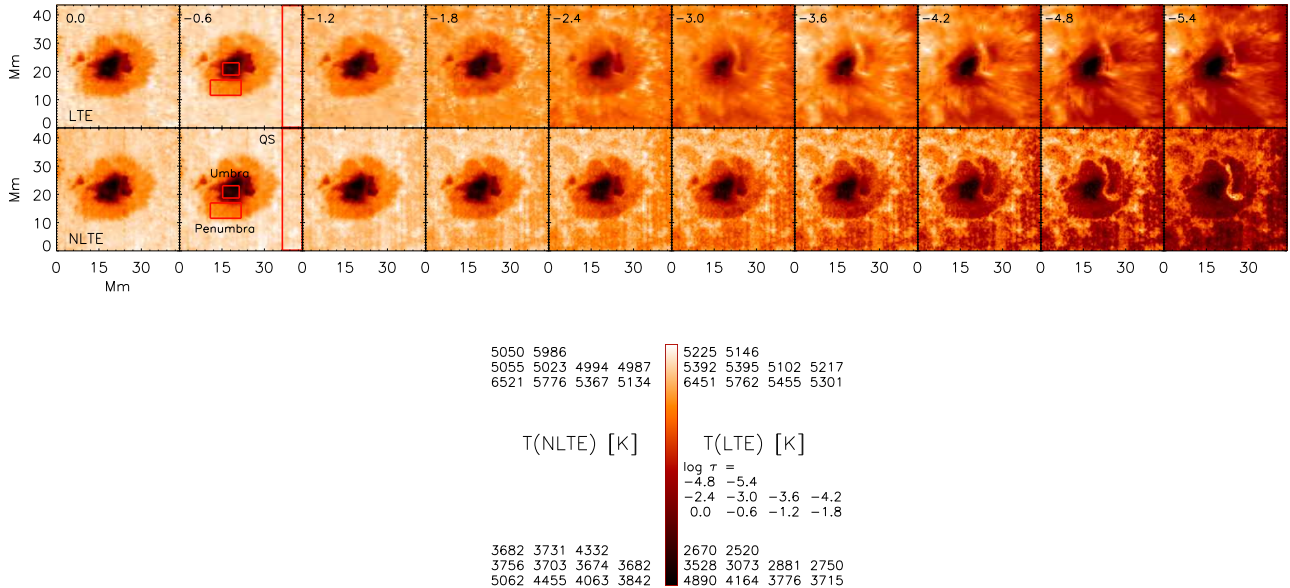}}}
\caption{Comparison of LTE (top row) and NLTE temperature (bottom row) at different levels of $\log\,\tau$. The color bar gives the range of displayed temperatures at the different $\log\,\tau$ levels. The red rectangles at $\log\,\tau = -0.6$ indicate the areas in umbra, penumbra and quiet Sun over which the ratio of the temperature in the LTE and NLTE inversion was calculated (cf.~Fig.~\ref{nlte_corr} below).\label{nlte_lte}}
\end{figure*}

\section{Results}
\subsection{LTE inversion}
Figures \ref{taumaps_ibis} and \ref{taumaps_spinor} show the LTE inversion results for the temperature at different optical depth layers for the IBIS and SPINOR data, respectively. The temperature contrast in the IBIS data is significantly larger compared to SPINOR because of the higher spatial resolution of the IBIS data. In both data sets, the LB stands out in the temperature maps at about $\log\,\tau \sim -3.0$. The smaller umbral core to the right of the LB is completely hidden under a series of filaments for layers above $\log \tau = -3.0$. The filamentary structure seen in the line-core image in Figure \ref{fov} is reproduced in the temperature maps at higher atmospheric layers. The inversion results from the IBIS spectra become slightly noisier at upper layers and more coarse at lower layers than the results derived from the SPINOR spectra. The former most likely is caused by the fact that the LTE inversion does not consider LOS velocities, which are larger in the IBIS data because of their higher spatial resolution, while the latter is caused by the coarser wavelength sampling in IBIS in the line wing (cf.~Figure \ref{IBIS_single_spec}) in comparison to the spectrograph data. To a first order, both 2D spectroscopic and spectrograph data yield comparable results. 

One of the main advantages of the LTE inversion over a more direct conversion from the intensity at a given wavelength to radiation temperature using the Planck function \citep[see, e.g.,][]{reardon+etal2009,beck+etal2012} -- apart from also taking the complete radiative transfer through the best-fit model atmosphere into account -- is that the resulting temperature stratifications provide a temperature cube vs.~optical depth. We converted each of the individual temperature stratifications from optical depth $\tau$ to geometrical height  $z$ using the tabulated relation between $\tau$ and $z$ as given in the HSRA. Three-dimensional (3D) renderings of the temperature cubes retrieved from the inversion are shown in Figure \ref{3d_spinor}, and also in animation 1 (LB in SPINOR data), animation 2 (full FOV in IBIS data) and animation 3 (loop structure in the moat) in the on-line section (see \citeauthor{beck+etal2014} \citeyear{beck+etal2014} for details of the generation of the animations). As shown in \citet{beck+etal2014}, the 3D renderings can be used to trace the thermal connectivity of individual features in the solar photosphere and chromosphere. The inversion of the complete time-series of IBIS scans allows one to trace the temporal evolution of the topology as well (animation 3).
\subsection{NLTE inversion and LTE temperature correction}
Figure \ref{nlte_lte} compares the results of the NLTE and LTE inversion of the SPINOR Ca 854.2\,nm spectra, while Figure \ref{scatter} below shows scatter-plots of the temperature derived by the two approaches for the same spectra. 
\begin{figure}
\centerline{\resizebox{8.5cm}{!}{\includegraphics{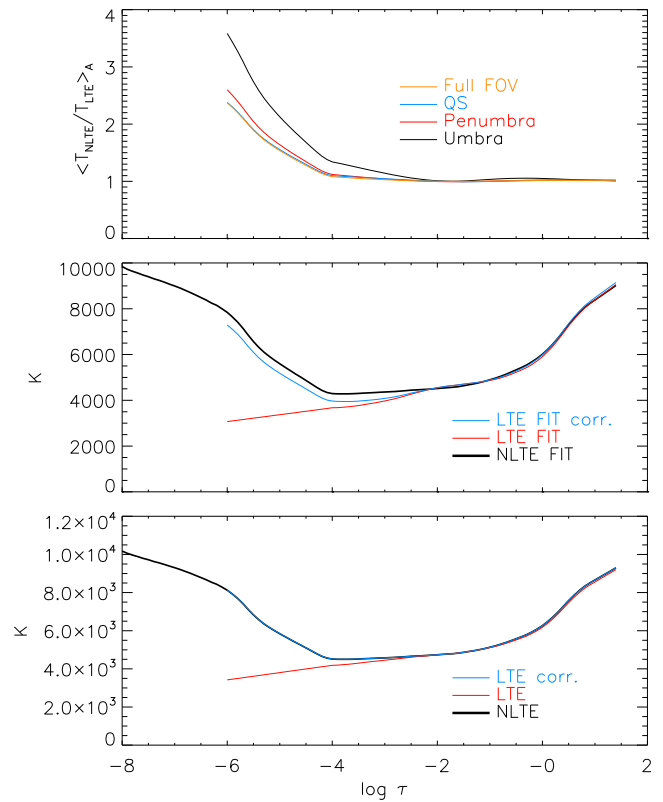}}}
\caption{Comparison of LTE and NLTE temperatures. Bottom panel: average temperature stratifications. Middle panel: temperature stratifications of a position in the penumbra. Top panel: NLTE temperature correction for the LTE results.\label{nlte_corr}}
\end{figure}
\begin{figure*}
\centerline{\resizebox{17cm}{!}{\includegraphics{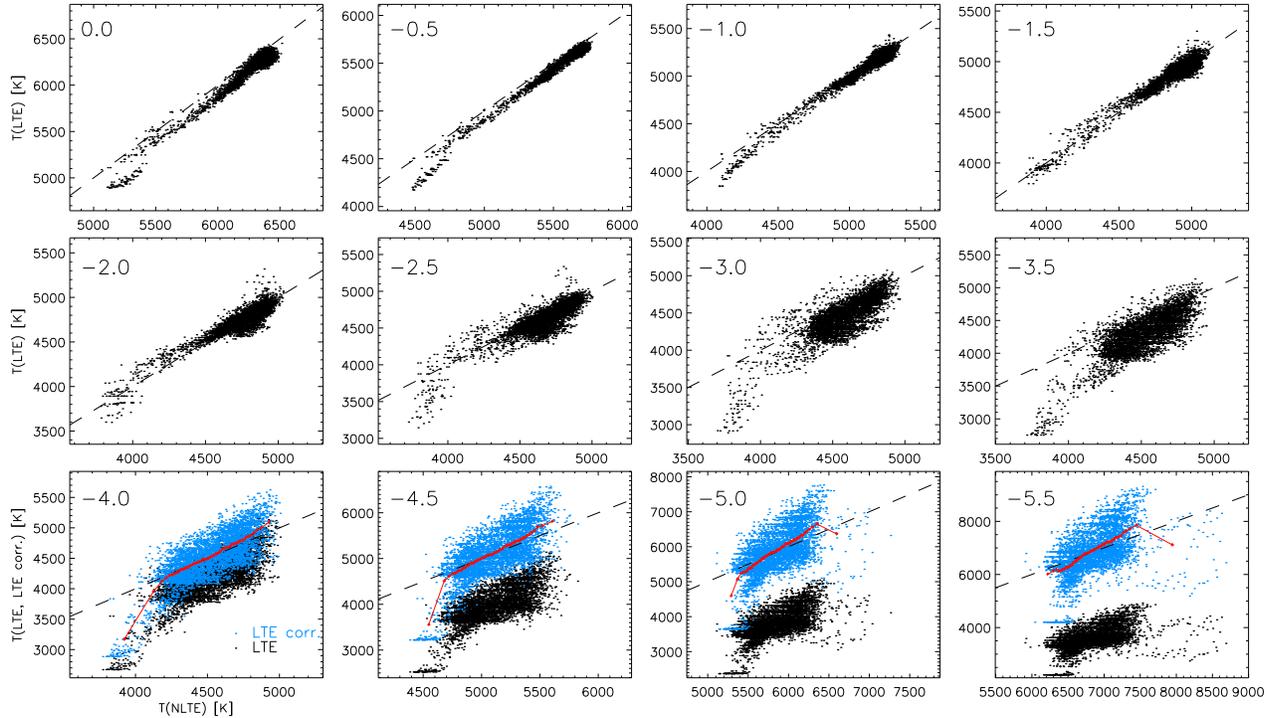}}}
\caption{Scatter-plots of the NLTE vs. LTE temperature with (blue dots) and w/o correction (black dots) in the full FOV. The black dashed lines indicate a one-to-one correlation. The red solid lines in the bottom row give the average value of the corrected LTE temperature (over-plotted in the bottom row as blue dots). The $\log\,\tau$ value is given in the upper left corner of each panel.\label{scatter}} 
\end{figure*}

The LTE and NLTE inversion results match quite close in the low atmosphere up to about $\log\,\tau \sim -1.8$ (Figure \ref{nlte_lte}). The spatial patterns in the LTE and NLTE temperature maps start to deviate higher in the atmosphere. The LTE temperature mimics the appearance of the spectra with increasingly more fibrils at upper layers corresponding to wavelengths close to the line core, while the LB becomes the most prominent feature in the NLTE results. 

The temperature in the LTE inversion does not increase in the upper atmosphere because of the limitation of the LTE assumption, while the NLTE inversion results shows the expected trend of increasing temperature (Figure \ref{nlte_corr}). This is especially clear in the average temperature stratifications shown in the bottom panel of Figure \ref{nlte_corr}. The scatter-plots in Figure \ref{scatter} reveal, however, that there is some correlation between the temperatures of the LTE and NLTE inversion even for $\log\,\tau < -3$. This was seen much clearer when only the uncorrected LTE-NLTE relation was plotted with its smaller temperature range on the $y$-axes of the bottom row of Figure \ref{scatter}. The main offset from the unity relation was found to be an optical-depth-dependent temperature offset rather than a lack of correlation. 

We therefore calculated the ratio between the individual temperature stratifications of the NLTE and LTE inversion results for each spatial position in four regions inside the FOV. We selected one area in the quiet Sun, in the penumbra and in the umbra (red rectangles at $\log \tau = -0.6$ in Figure \ref{nlte_lte}) and also averaged the resulting ratio of temperatures over the entire FOV. The ratios of NLTE to LTE temperatures (top panel of Figure \ref{nlte_corr}) stay close to unity up to about $\log \tau = -3.0$, but increase to a factor of about 2.5 at $\log \tau = -6.0$ in the quiet Sun and to a factor of about 4  at $\log \tau = -6.0$ in the umbra. 

The temperature ratio provides a basic correction factor for the temperature stratifications derived by the LTE inversion. Applying the correction given by the average of the ratio over the full FOV separately to all individual LTE temperature stratifications and calculating again the average temperature stratification after the correction yields naturally a close match to the average temperature stratification of the NLTE inversion (blue and black line in bottom panel of Figure \ref{nlte_corr}). Using a randomly picked position inside the penumbra shows that the correction only converts nearly perfectly from LTE to NLTE on average, but it still reduces the mismatch between the LTE and NLTE approach significantly (blue and black line in the middle panel of Figure \ref{nlte_corr}). The scatter plots of Figure \ref{scatter} show the same effect of matching LTE to NLTE temperatures by the correction factor. The average values of LTE and NLTE temperatures above $\log \tau = -4.0$ (red lines in the bottom row of Figure \ref{scatter}) follow a one-to-one correlation, with a scatter that is comparable to the uncorrected LTE case.

\section{Summary and Discussion}
We presented a fast ($\ll 1$\,s per profile) inversion code for \ion{Ca}{ii} IR spectra at 854.2\,nm. We found no significant difference when applying it to moderately sparsely sampled 2D spectroscopic data or spectrograph data with a larger spectral coverage and equidistant spectral sampling. The inversion code delivers temperature stratifications assuming LTE. From a comparison with a NLTE inversion of the same spectra, we derived a correction curve for the LTE temperature stratifications that on average matches LTE and NLTE results.

The archive of temperature stratifications that is part of the code can be directly used to synthesize other spectral lines as well, e.g., the other \ion{Ca}{ii} IR lines at 849.8 and 866.2\,nm, or \ion{Ca}{ii} H and K. The correction of the LTE temperatures to match more closely the more sophisticated NLTE result should not depend strongly on which spectral line is actually inverted. There should, however, be an additional deviation between LTE and NLTE temperatures for \ion{Ca}{ii} H and K. Partial frequency redistribution is more important for these lines than for the \ion{Ca}{ii} IR lines, while both our LTE code and the current version of NICOLE assume complete frequency redistribution \citep{socasnavarro+etal2014}.

A replacement of the LTE archive with any suited set of NLTE spectra would be the best possible improvement. This should be possible with the release of NICOLE or using the RH code \citep{uitenbroek2001}. The inversion code is still lacking a good method to include LOS velocities in the approach. The line cores of all \ion{Ca}{ii} lines can exhibit intensity reversals that prevent using the standard velocity proxies used with photospheric lines. For most of the data that we inverted up to now, the lack of LOS velocities was, however, not critical for the inversion. At the typical spectral sampling of the data, even Doppler shifts from sonic flows yield only a displacement of less than 5 pixels. Without velocity gradients, the LTE archive profiles are always perfectly symmetric around the line core. Another feature not taken into account currently is the inclination of the LOS to the solar surface vertical for off-centre observations. This presumably would require to synthesize the archive spectra taking the heliocentric angle of the observations into account. 

The current inversion code can be applied to any kind of Ca 854.2\,nm spectra, with the caveat that an even more sparse sampling than in the current case 
(27 wavelength points) might turn out to be critical. With its speed, the code allows us to invert even series of, e.g., IBIS spectra with about 700,000 profiles per spectral scan in a reasonable time. It might be possible to speed up the inversion by applying a principal component analysis to the synthetic archive spectra \citep[cf.][]{casini+etal2013}. It would be potentially interesting to apply it routinely to the full-disk Ca 854.2\,nm spectra taken by the Synoptic Optical Long-term Investigations of the Sun \citep[SOLIS;][]{keller+etal2001,raouafi+etal2008} telescope.
\begin{figure*}
\centerline{\resizebox{17cm}{!}{\includegraphics{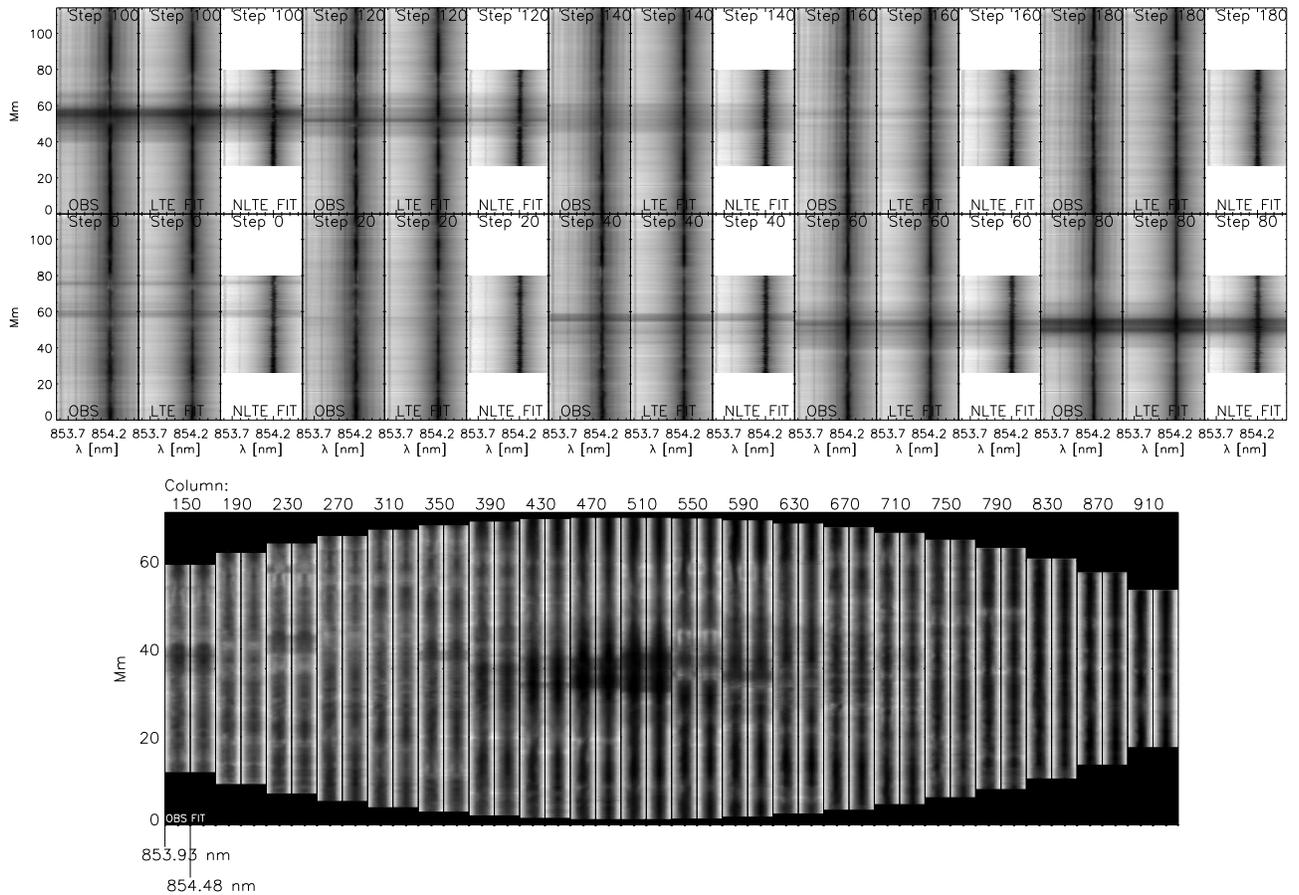}}}
\caption{Comparison of observed and best-fit spectra. Top panel: SPINOR slit spectra. Bottom panel: spectra along columns of the IBIS FOV. For the SPINOR data, the left sub-panel always shows the observed spectra, the middle sub-panel the LTE best-fit spectra and the right sub-panel the NLTE best-fit spectra. For the IBIS data, only observed and LTE best-fit spectra are shown. \label{slit_spec_spinor}}
\end{figure*}
\section{Conclusions}
Our inversion approach provides a method for a fast analysis of chromospheric \ion{Ca}{ii} spectra to derive an estimate for the corresponding temperature stratifications. The results can be roughly corrected for NLTE effects using a static temperature correction curve derived from a comparison to an NLTE inversion. Thanks to its speed, an application to even extended data sets of millions of spectra, like nowadays obtained routinely with modern fast 2D spectrometers, is possible in a reasonable time without resorting to massive computations.
\begin{acknowledgements}
The Dunn Solar Telescope at Sacramento Peak/NM is operated by the National Solar Observatory (NSO). NSO is operated by the Association of Universities for Research in Astronomy (AURA), Inc.~under cooperative agreement with the National Science Foundation (NSF). R.R. and  R.E.L. acknowledge financial support by the DFG grants RE 3282/1-1 and DFG 787/3-1, respectively. We thank especially H. Socas-Navarro (IAC) for providing us with the NICOLE code.
\end{acknowledgements}
\bibliographystyle{aa}
\bibliography{references_luis_mod}

\begin{appendix}
\section{Fit quality}\label{fit_quality}
\begin{figure*}
\centerline{\resizebox{17cm}{!}{\includegraphics{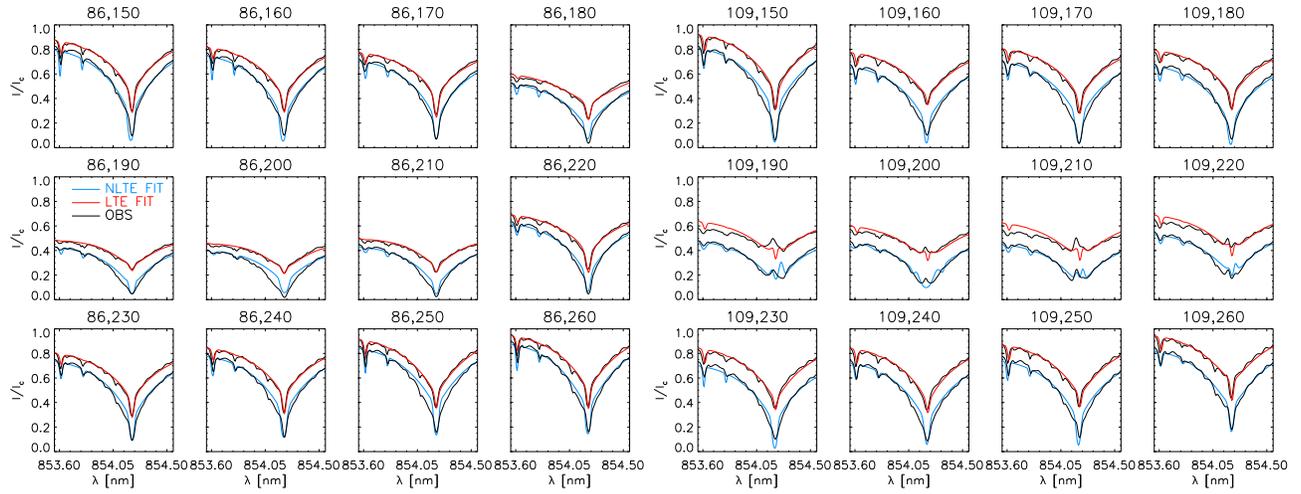}}}
\caption{Individual spectra in the SPINOR data on vertical cuts across the sunspot. Left twelve panels: at scan step 86. Right twelve panels: at scan step 109 across the LB. Black/red lines at top in each panel: observed/best-fit spectra in the LTE fit. Black/blue lines at bottom: observed/best-fit spectra in the NLTE fit displaced by an offset in y for better visibility.  \label{single_spec}}
\end{figure*}

\begin{figure*}
\centerline{\resizebox{17cm}{!}{\includegraphics{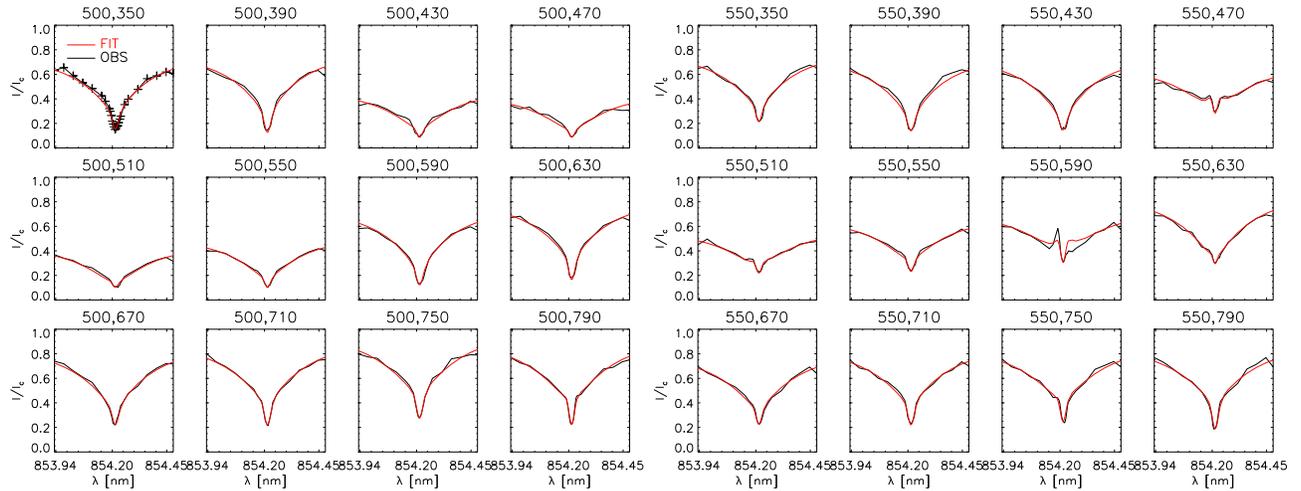}}}
\caption{Individual spectra in the IBIS data on vertical cuts across the sunspot. Left panels: at column 500. Right panel: at column 550 across the LB. Black/red lines: observed/best-fit spectra. The pluses in the upper left panel denote the sampled wavelengths. \label{IBIS_single_spec}}
\end{figure*}

Figures \ref{slit_spec_spinor} to \ref{IBIS_single_spec} show comparisons of observed and best-fit spectra. The top panel of Figure \ref{slit_spec_spinor} shows slit-spectra of SPINOR for 10 scan steps with the corresponding LTE and NLTE best-fit spectra while the  bottom panel shows 20 sets of spectra along columns of the IBIS FOV. In the SPINOR best-fit spectra, the vertical interference fringes are missing because the inversion codes have no degree of freedom to reproduce them. For most spectra neglecting the LOS velocities is seen to have no real impact on the fit quality, especially in the line wings. The same generally good match of observed and best-fit spectra is seen for the IBIS spectra in the bottom panel of Figure \ref{slit_spec_spinor}.

The individual spectra in Figures \ref{single_spec} and \ref{IBIS_single_spec} show for which profiles the LTE fit procedure fails and where also the NLTE inversion encounters difficulties. Note that the observed spectra used as input for the NLTE fit were not subjected to the same series of correction steps as needed for the LTE fit and thus differ slightly in, e.g., the line depth or the shape of the line close to the line core. In the left halves of both figures spectra along a cut across the sunspot are shown. These observed spectra are usually well reproduced by the LTE and NLTE best-fit spectra. 

The middle panels of the right halves of Figures \ref{single_spec} and \ref{IBIS_single_spec} correspond to where the cut passed across the location of the light bridge. Here, the inversion approach is often not able to reproduce the observed shape with a single, large emission peak. Such profiles are not contained in the LTE archive, thus only the line wing can be fully fitted. In the line core sometimes only a rough match is achieved in the LTE case by using a profile with a double-peaked emission pattern that mimics the line-core intensity of the observed profile but not its actual shape (cf.~the middle right panels of Figure \ref{single_spec}). The (one-component) NLTE inversion also had problems with these ``abnormal'' profiles in the light bridge. On a few test inversions, the final NLTE best-fit solution would vary between large red and blue shifts with a corresponding increase and decrease of temperature, to create a single emission peak either by strong Doppler-shifted emission or strong Doppler-shifted absorption. For such profiles, an explicit modeling of the stratification might be required, at least as initial model, because even the NLTE fit could not directly find a good or unique solution for the atmospheric stratification.
\end{appendix}

\end{document}